\documentclass[fleqn,twoside,epsf]{article}
\usepackage{espcrc2}

\usepackage{graphicx}

\newcommand{\AmS}{{\protect\the\textfont2
  A\kern-.1667em\lower.5ex\hbox{M}\kern-.125emS}}

\newcommand{\be}{\begin{equation}}
\newcommand{\ee}{\end{equation}}
\newcommand{\beq}{\begin{eqnarray}}
\newcommand{\eeq}{\end{eqnarray}}

\title{Hadron deformation from Lattice QCD~\thanks{This research is supported in part by funds provided by the Levendis Foundation}}

\author{C. Alexandrou\address{Department of Physics, University of Cyprus,
CY-1678 Nicosia, Cyprus}}

\begin{document}

\begin{abstract}
We address the issue of hadron deformation within the framework of lattice
QCD. For hadrons with spin greater than 1/2 the deformation can be determined
 by evaluating the charge and matter distributions.
Deviation of the nucleon shape from spherical symmetry is
determined by evaluating the quadrupole strength in the transition
$\gamma N  \rightarrow \Delta(1232)$, both in the quenched and
in the unquenched theory.
\vspace{1pc}
\end{abstract}

\maketitle

\vspace*{-1.75cm}

\section{Introduction}

Hadronic matrix elements of two- and three- current correlators 
encode detailed information on hadron structure
such as
quark spatial distributions, charge radii and quadrupole moments.
 In the non-relativistic limit they
 reduce to the square of the wave function and 
therefore the  shape of the hadron can be extracted. 
The issue of a deformation in the nucleon
and $\Delta$ was first examined in the context of the constituent
quark model~\cite{Isgur} with the one gluon exchange giving rise to 
a colour magnetic dipole-dipole interaction. The tensor  component
of this interaction
causes a D-wave admixture in the nucleon and $\Delta$ wave function
and leads to a deformation.
Other mechanisms have been invoked to explain  hadron deformation:
In cloudy bag models the deformation is caused by the asymmetric pion
 cloud~\cite{Eisenberg}. 
In soliton
models it is thought to be due to the  non-linear pion field interaction.
In the constituent quark model it was recently proposed to be
due to two-body contributions
to the electromagnetic current arising from  the  elimination of gluonic 
and quark-antiquark degrees of freedom~\cite{Buchmann}. 
Because of the strong experimental indications 
that the nucleon is deformed,
it is interesting to address the general 
issue of hadron deformation within
lattice QCD and try to understand its physical origin. 
In the first part of this talk we present results
on  
the charge and matter density distributions
 evaluated on the lattice, both in the quenched and in the unquenched theory.
A comparison between the charge and matter distributions suggests that
deformation can arise from the relativistic motion of the quarks. 

For the nucleon, being a spin 1/2 particle,
the spectroscopic quadrupole moment averages to zero 
and therefore it cannot be studied via the density distributions.
For this  reason we 
look for a quadrupole strength
in the  $\gamma N \> \rightarrow \> \Delta$ transition
as is done in experimental studies.
 Spin-parity selection rules allow a magnetic dipole, M1, an electric
quadrupole, E2, or a Coulomb quadrupole, C2, amplitude for this transition.
If both the nucleon and the $\Delta$
are spherical, then E2 and C2 
should be zero. Although M1 is indeed the dominant amplitude,
there is mounting experimental evidence over  a range of momentum transfers
that E2 and C2 are
 non-zero~\cite{Bates}.
State-of-the-art lattice QCD calculations can yield model independent
results on these form factors and provide direct comparison with experiment.
The second part of this talk discusses  the evaluation of the Sachs transition
form factors ${\cal G}_{M1}$, ${\cal G}_{E2}$ and ${\cal G}_{C2}$~\cite{Jones}
 describing $\gamma N \rightarrow \Delta$.

\vspace*{-0.2cm}

\section{Hadron wave functions}

\vspace*{-0.2cm}

\begin{figure}[h]
\vspace*{-0.75cm}
\tabskip=0pt\halign to\hsize{\hfil#\hfil\tabskip=0pt plus1em&
\hfil#\hfil\tabskip=0pt\cr
\hspace*{-0.5cm}\mbox{\includegraphics[height=3.cm,width=4cm]{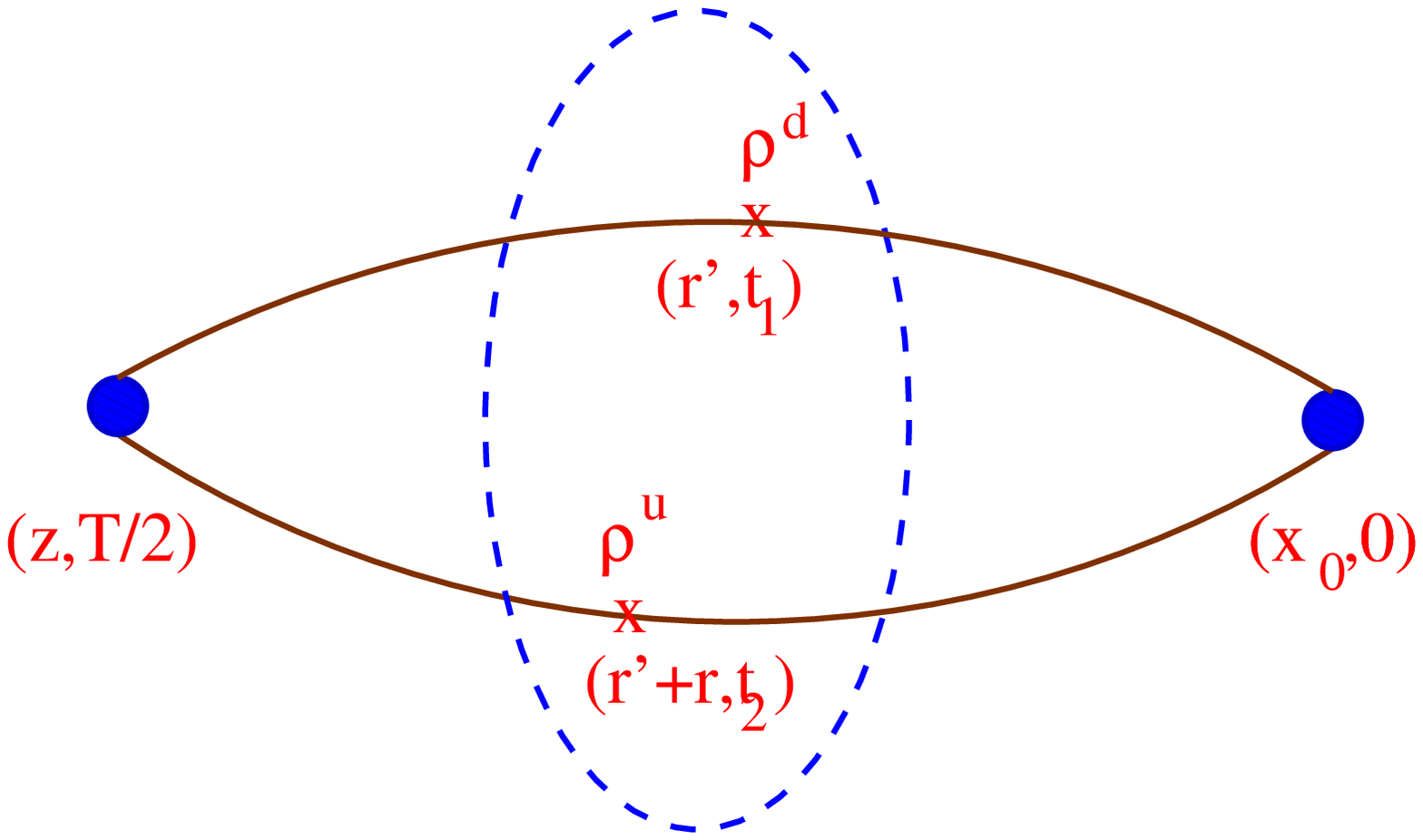}}
& \hspace*{0.1cm}
\mbox{\includegraphics[height=5cm,width=4cm]{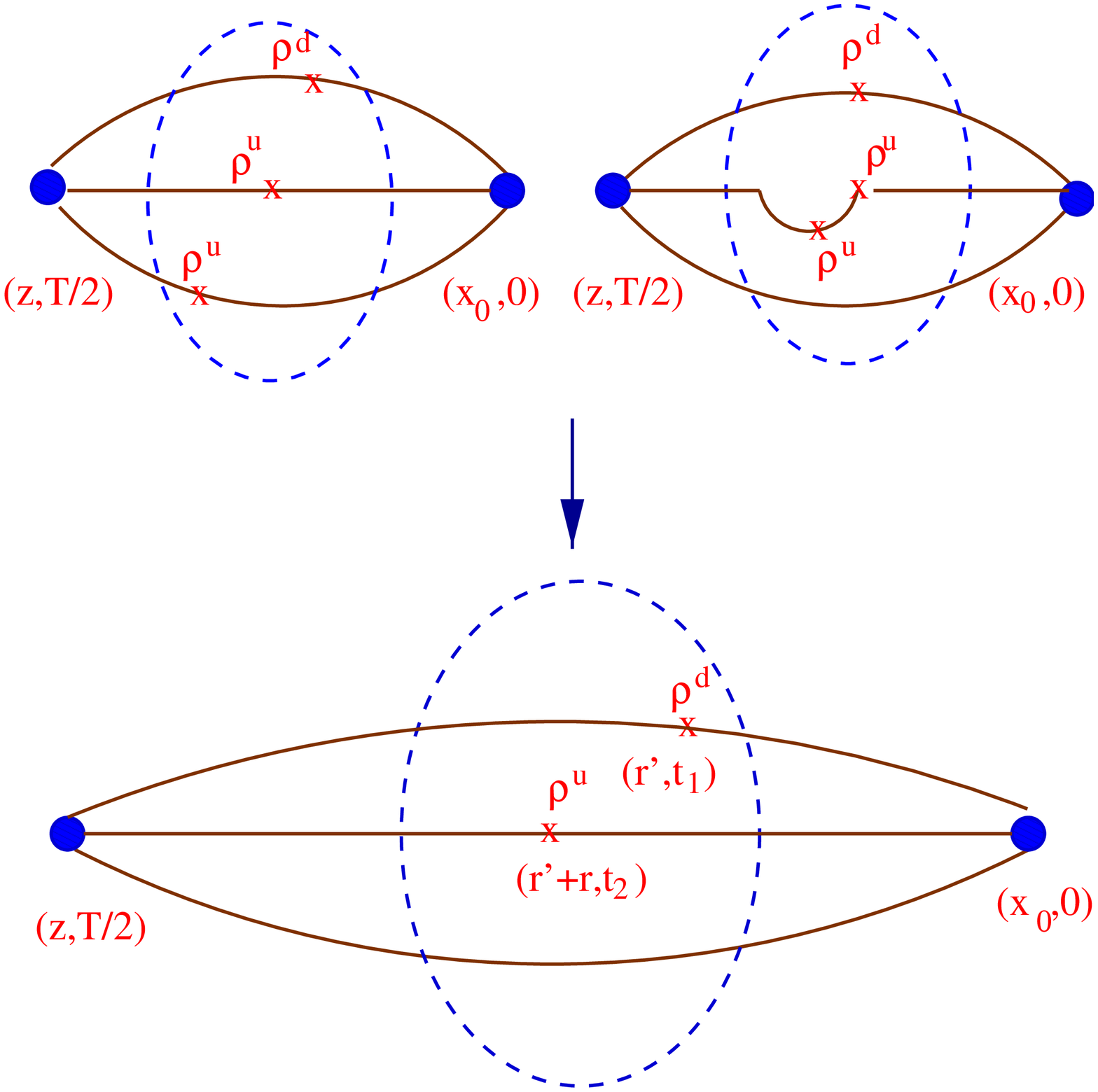}}
\cr
Meson & Baryon \cr}
\vspace*{-0.8cm}
\caption{Two- and three-  current correlators for a meson (left)
and a baryon (upper right); Lower right: two-current correlator
for a baryon after integration over one relative distance.
These diagrams are drawn 
assuming anti-periodic boundary conditions in the time 
direction.}
\vspace*{-0.7cm}
\label{fig:dd2}
\end{figure}

For a meson we evaluate  the two-current correlator
shown schematically in Fig.~\ref{fig:dd2}. Taking the current insertions at equal
times $t_1=t_2=t$ this correlator is given by  

\vspace*{-0.2cm}

\be
C_\Gamma({\bf r},t) = \int\> d^3r'\>
\langle h|j_\Gamma^{u}({\bf r}'+{\bf r},t)j_\Gamma^{d}({\bf r}',t)|h\rangle
\label{correlator}
\ee
for a hadron state $|h\rangle$. 
The current operator  is given by the normal ordered product

\vspace*{-0.4cm}

$$
j_\Gamma^u({\bf r},t) = :\bar{u}({\bf r},t)\Gamma u({\bf r},t):
\hspace*{0.3cm} {\rm where} 
\hspace*{0.3cm}\Gamma=\left\{ \begin{array}{c}
\gamma_0\\  1  \end{array} \right.
$$

\noindent
for the charge and matter correlators respectively.
We note that no gauge ambiguity arises in
this determination of hadron wave functions, 
unlike Bethe-Salpeter amplitudes.
In the case of  baryons two relative distances are
involved and three current insertions
are required as shown in Fig~\ref{fig:dd2}. 
However we may consider integrating over one relative 
distance
to obtain the one-particle density distribution that involves two current
insertions 
 and is thus evaluated via Eq.~\ref{correlator} for $t_1=t_2=t$.

All the results on the current-current correlators
 have been obtained on  lattices of size
$16^3 \times 32$. We analyse, for the quenched case, 220 NERSC 
configurations  at $\beta=6.0$ and, for the unquenched,
150 SESAM configurations~\cite{SESAM} for $\kappa=0.156$ and 200 
for $\kappa=0.157$
simulated at $\beta=5.6$ with two dynamical quark flavours.
The physical volume of the quenched and unquenched lattices
is approximately the same.

\begin{figure}[h]
\begin{center}
\mbox{\includegraphics[height=5cm,width=6cm]{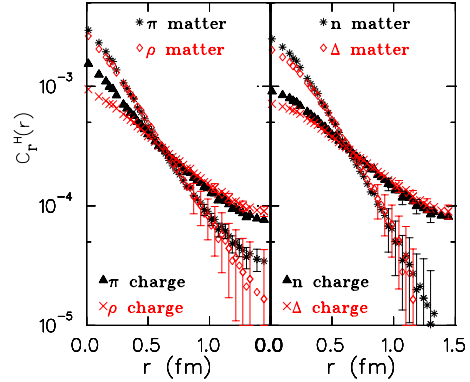}}
\vspace*{-1cm}
\end{center}
\caption{Quenched charge  and matter 
 distributions, $C_{\gamma_0}(r)$ and
 $C_I(r)$, at $\kappa=0.153$. Left for the pion and the rho;
Right for the nucleon and the $\Delta$.}
\vspace*{-0.8cm}
\label{fig:ch_matter}
\end{figure}

In Eq.~\ref{correlator} the current couples to the quark at fixed time separation $t$ from the source,
which must be  large enough to 
sufficiently isolate the hadron ground state.
We checked, by comparing data  at different $t$, 
that excited states 
are sufficiently suppressed at 
maximal  time separation $t=T/4$ 
in  the current-current correlators~\cite{AFT2} where
we have taken anti-periodic boundary conditions in the temporal direction.
This
allows us to analyse standard full QCD
configurations that
employ anti - periodic boundary conditions.
Since, for our
parameters,   local sources produce the same results for $C_\Gamma(r)$ 
as smeared ones and carry no gauge noise, we use them
in this study.

The quenched charge and matter density distributions 
were studied as a function of  the naive quark mass, 
$ m_q = 1/2(1/\kappa-1/\kappa_c)$, where $\kappa_c$ is the
hopping parameter value at which the pion becomes massless. 
The dependence 
 of both distributions on $m_q$
is more pronounced for the mesons than for the baryons.  
In particular the pion charge and matter distributions
 broaden as the quarks become lighter~\cite{AFT,AFT2}.
In the case of the rho the charge distribution shows 
a stronger dependence on the quark mass than  the matter distribution,
for which  no further mass dependence is observed for 
  $m_q \stackrel{<}{\sim} 200$~MeV. 
For  the nucleon and the $\Delta^+$ 
essentially no variation is seen
over  the range of naive quark masses $\sim  300-100$~MeV
investigated here. 
The quenched  charge and matter  distributions are compared in 
Fig.~\ref{fig:ch_matter} at $\kappa=0.153$. The matter distribution
is for all hadrons less broad than the charge distribution.
The root mean square radius (rms) can be extracted
 by evaluating

\be
\hspace*{2.cm} \langle r^2\rangle = \frac{\sum_{\bf r} {\bf r}^2 C_\Gamma({\bf r}, t)}
                          {\sum_{\bf r} C_\Gamma({\bf r}, t)} \quad.
\ee
For non-relativistic states the charge rms radius  can be written in terms of
the form factors as~\cite{Negele}

\vspace*{-0.4cm}
\beq
  \langle r^2\rangle &=&-\sum_n \nabla^2_{{\bf q}^2} F_{hn}^u({\bf q})
                            F_{nh}^d(-{\bf q})\left.\right |_{{\bf q}^2=0} 
\nonumber \\
&=&  r_u^2+r_d^2 -2 \sum_{n}{\bf d}_{hn}^u.{\bf d}_{nh}^d
\label{rms}
\vspace*{-0.5cm}
\eeq
where the non-relativistic form factors
$F_{hn}^{u,d}({\bf q})=\langle h,{\bf q} |j_{\gamma_0}^{u,d}|n,{\bf 0} \rangle$
 are given by
\be
F_{hn}^{u,d}({\bf q})=\left\{\begin{array}{cc} 
1-r^2_{u,d}\> {\bf q}^2/6 \> +{\cal O}({\bf q}^4) & n=h \\
{\bf q}.{\bf d}_{hn}^{u,d} +  {\cal O}({\bf q}^2) & n \neq h\end{array} \right.
\ee
for u- and d- quarks.
The dipole-dipole term appearing in Eq.~\ref{rms} can be evaluated 
 if, for instance, we assume a non-relativistic two-body system with equal
masses for the u and d
quarks in the center of mass frame, since then ${\bf r}_u=-{\bf r}_d$ and $<r^2>=2(r_u^2+r_d^2)$.
Only if we allow current insertions at unequal times $t_1$ and $t_2$
and take $t_2-t_1$ large enough so that intermediate excited states
are sufficiently suppressed then the charge radius, for instance of the pion, 
 is obtained from

\be
\left.\frac{\partial C_{\gamma_0}({\bf q})}{\partial {\bf q}^2} 
\right |_{{\bf q}^2=0} = \left.2\frac{\partial F_{\pi}({\bf q})}
{\partial {\bf q}^2}\right |_{{\bf q}^2=0} = -\frac{2}{6}\langle r^2_{\rm ch}\rangle\quad .
\ee

\begin{figure}[h]
\vspace*{-0.5cm}
\begin{center}
\mbox{\includegraphics[height=4.5cm,width=6cm]{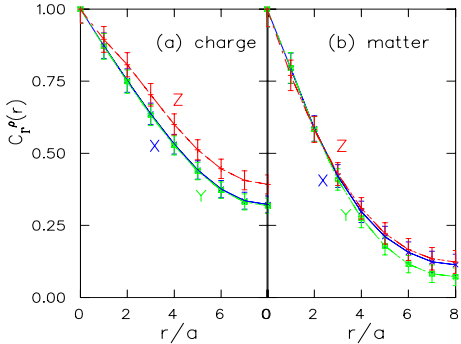}}
\vspace*{-1.cm}
\end{center}
\caption{$C_\Gamma(0,0,z)$ (Z-curve);
$C_\Gamma(x,0,0)$ and $C_\Gamma(0,y,0)$ ($X,Y$-curves).}
\vspace*{-0.75cm}
\label{rho-asymmetry}
\end{figure}

\begin{figure}[h]
\begin{center}
\mbox{\includegraphics[height=6cm,width=6cm]{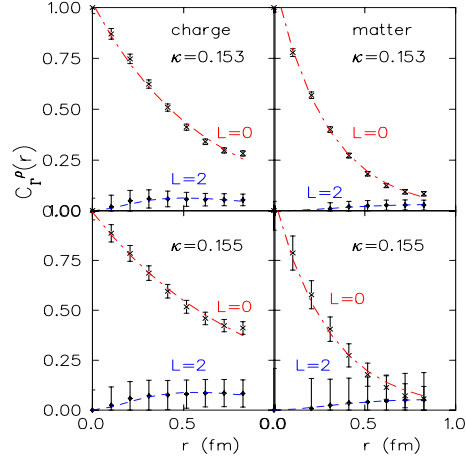}}
\end{center}
\vspace*{-1.cm}
\caption{
 Decomposition into 
 $L=0$ and $L=2$ contributions 
at $\kappa=0.153$  and $\kappa=0.155$.
Left: charge distribution, right: matter distribution.}
\label{deformation}
\vspace*{-0.75cm}
\end{figure}

The rho charge density distribution~\cite{AFT,panic02}
produces a non-zero quadrupole moment
or equivalently a $z-x$ asymmetry as shown in Fig.~\ref{rho-asymmetry},
where the $z$-axis is taken along
the spin of the rho.
This yields a  deformation  
$\delta\equiv \frac{3}{4}\> \frac{\langle 3z^2-r^2 \rangle}{\langle r^2 \rangle}\sim  0.03\pm 0.01$ in the chiral limit.
 An angular decomposition 
 of the wave function, as shown in Fig.~\ref{deformation}, corroborates
 a non-zero charge deformation by
 producing a non-zero $L=2$ component~\cite{Gupta}.
No such  deformation is 
seen  for the matter density distribution in Figs.~\ref{rho-asymmetry}
and \ref{deformation} .
 Since the matter and charge operators
have the same  non-relativistic limit,
this suggests  that hadron
charge deformation is a relativistic effect.
This result has
 strong implications for the validity of various phenomenological models used
in the study of nucleon deformation.

\begin{figure}[h]
\vspace*{-0.5cm}
\begin{center}
\mbox{\includegraphics[height=5cm,width=6cm]{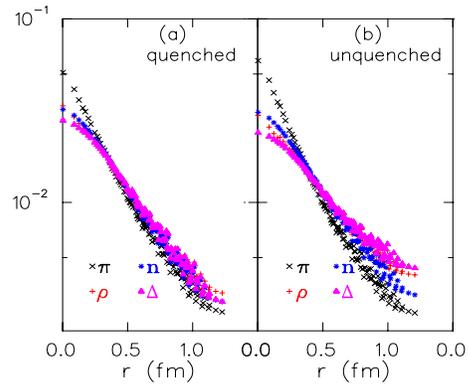}}
\vspace*{-1cm}
\end{center}
\caption{Quenched  and unquenched  charge density  distributions
 at $\kappa=0.153$  and $\kappa=0.156$ respectively.
 Errors bars are omitted for clarity.}
\label{fig:charge_full}
\end{figure}

\begin{figure}[h]
\begin{center}
\vspace*{-0.5cm}
\mbox{\includegraphics[height=5cm,width=6cm]{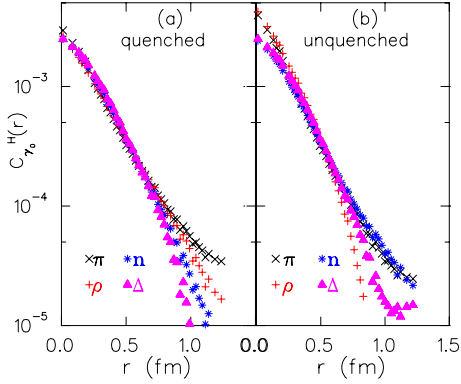}}
\vspace*{-1cm}
\end{center}
\caption{Quenched  and  unquenched  matter density  distributions
 at $\kappa=0.153$  and $\kappa=0.156$ respectively.
Errors bars are omitted for clarity.} 
\vspace*{-0.75cm}
\label{fig:matter_full}
\end{figure}

\begin{figure}[h]
\begin{center}
\vspace*{-0.7cm}
\mbox{\includegraphics[height=4cm,width=4cm]{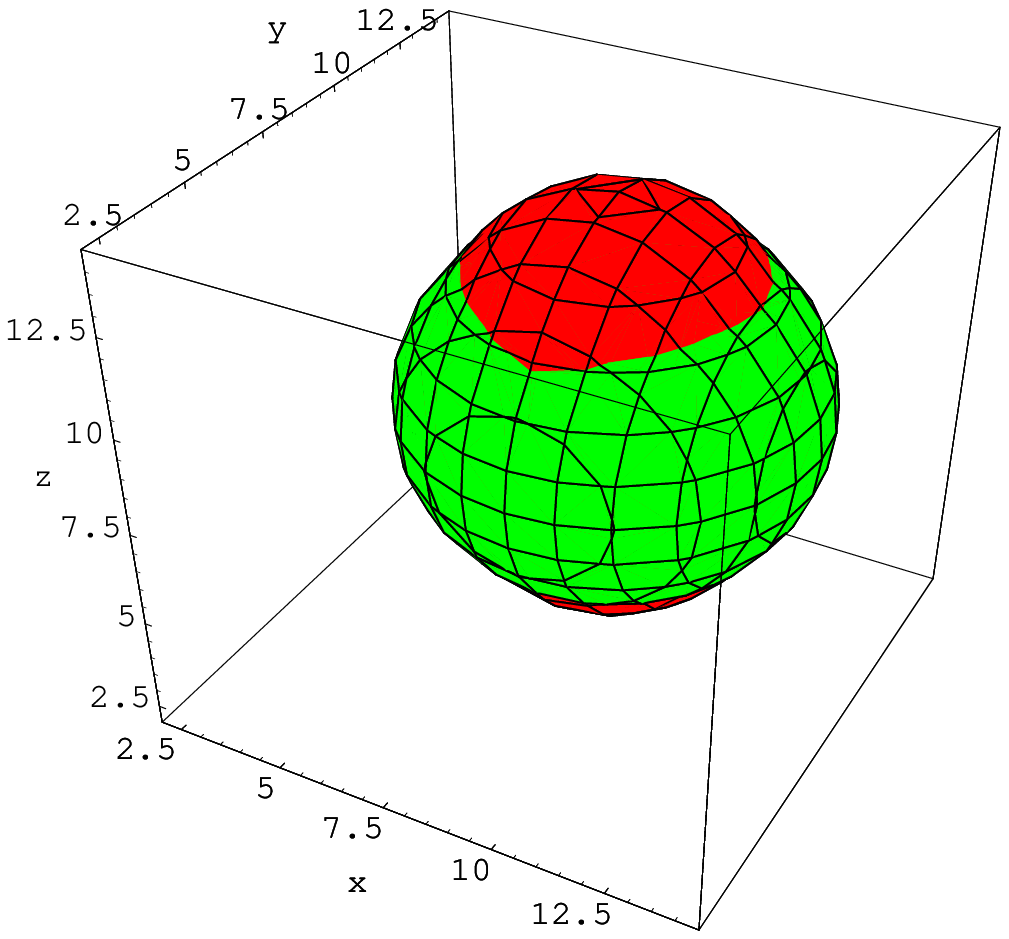}}
\mbox{\includegraphics[height=4cm,width=4cm]{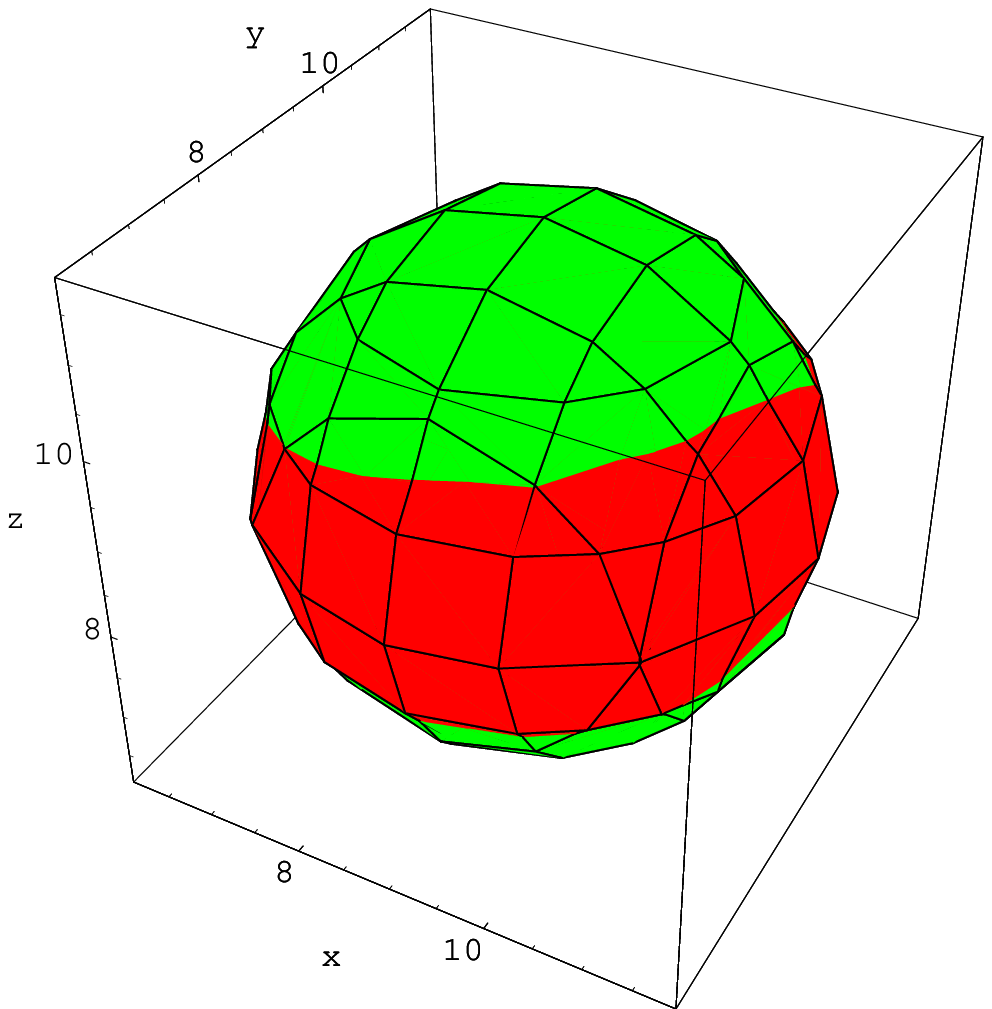}}
\vspace*{-1cm}
\end{center}
\caption{Three-dimensional contour plot of the charge correlator (red):
upper for the rho state  (cigar shape)
 and  lower for the $\Delta^+$  (slightly oblate)
for two dynamical quarks at $\kappa=0.156$.
Values of the correlator (0.5 for the rho, 0.8 for the $\Delta^+$) were chosen
to show large distances but avoid finite-size effects.
We have included for comparison
 the contour of a sphere (green).}
\vspace*{-0.7cm}
\label{fig:3-d}
\end{figure}

Unquenched results obtained at 
$\kappa=0.156$ $(m_\pi/m_\rho=0.83)$ and 
$\kappa = 0.157$ $(m_\pi/m_\rho=0.76)$
can be compared  to quenched ones at $\kappa=0.153$ $(m_\pi/m_\rho=0.84)$ and 
$\kappa=0.154$ $(m_\pi/m_\rho=0.78)$ having
similar pion to  rho mass ratios.  
In  Figs.~\ref{fig:charge_full} and \ref{fig:matter_full}
unquenched  results for $C_{\gamma_0}(r)$ and   $C_I(r)$ at $\kappa=0.156$ 
are compared to quenched results at $\kappa=0.153$. 
The unquenched charge  and matter distributions
show an increase at short
distances in the case of the pion and the rho whereas for the baryons
 no significant changes are seen.
Like in the quenched case the  unquenched matter distribution
shows a faster fall-off as compared to that
observed for the charge
distribution.
Whereas unquenching
leads to an increase in the  rho charge asymmetry
and to a small deformation for the $\Delta^+$, as shown by the
3-dimensional contour plots of Fig.~\ref{fig:3-d}, 
it has no effect on the matter density distribution.
This again suggests a relativistic origin for the rho charge deformation. 
 Pion cloud contributions
to hadron deformation are
 expected to become significant
as we approach the physical pion mass and 
a lattice calculation with lighter pions
can test if the charge rho asymmetry shows
a significant increase,  as suggested
in some models, and at the same time if a matter density asymmetry shows up.

\section{$\gamma N \rightarrow \Delta$ transition form factors}

\begin{figure}[h]
\begin{center}
\mbox{\includegraphics[height=3cm,width=5cm]{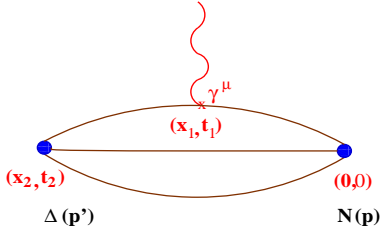}}
\vspace*{-0.8cm}
\end{center}
\caption{$\gamma N \rightarrow \Delta$ matrix element. The photon couples
to a quark in the nucleon
  at time separation $t_1$ from the source to produce a $\Delta$.}
\label{NDelta}
\vspace*{-0.8cm}
\end{figure}

We evaluate the 3-point function,
$\langle G^{\Delta j^\mu N}_{\sigma} (t_2, t_1 ;{\bf p}^{\;\prime}, {\bf p} ; \Gamma) \rangle$
shown schematically in Fig.~\ref{NDelta}, as well as
 $\langle G^{N j^\mu \Delta}_{\sigma}
(t_2, t_1 ; {\bf p}^{\;\prime}, {\bf p}; \Gamma) \rangle$ for the
reverse $\Delta \rightarrow \gamma N$ transition.
Exponential decays and normalization constants cancel in the ratio
$R_\sigma (t_2, t_1; {\bf p}^{\; \prime}, {\bf p}\; ; \Gamma ; \mu)=$

\vspace{-0.4cm}

\beq
& \>&\hspace*{-1.2cm}\large{
 \left [\frac{
\langle G^{\Delta j^\mu N}_{\sigma} (t_2, t_1 ; {\bf p}^{\;\prime}, {\bf p};
\Gamma ) \rangle \;
\langle G^{N j^\mu \Delta}_{\sigma} (t_2, t_1 ; -{\bf p}, -{\bf p}^{\;\prime};
\Gamma^\dagger ) \rangle }
{
\langle \delta_{ij} G^{\Delta \Delta}_{ij}(t_2,{\bf p}^{\; \prime};
\Gamma_4) \rangle \;
\langle G^{NN} (t_2, -{\bf p} ; \Gamma_4) \rangle } \right]^{1/2}
} \nonumber \\
&\;&\hspace*{-.5cm}\stackrel{t_2 -t_1 \gg 1, t_1 \gg 1}{\Rightarrow}
\Pi_{\sigma}({\bf p}^{\; \prime}, {\bf p}\; ; \Gamma ; \mu) \; ,
\hspace*{2cm} (6) \nonumber
\label{R-ratio}
\eeq
where $ G^{NN}$ and $ G^{\Delta \Delta}_{ij}$ are the
 nucleon and $\Delta$ two-point functions evaluated in the standard way.

\stepcounter{equation}

\noindent
The Sachs form factors are obtained by appropriate combinations
of the $\Delta$ spin-index $\sigma$, current direction $\mu$
and projection matrices $\Gamma$. For instance,
in the
$\Delta$ rest frame $ \vec{p}^{\;\prime} = 0 \;, \; \vec{q} = (q,0,0)$
~\cite{Leinweber,All} we have

\vspace*{-0.3cm}

$$
 \hspace*{-2.5cm}{\cal G}_{M1} = {\cal A}\;\frac{1}{|{\bf q} |} \;
\Pi_2({\bf 0}, {\bf -q}\; ; +i \;\Gamma_4 ; 3)
$$

\vspace*{-0.5cm}



$$
{\cal G}_{E2} = {\cal A} \; \frac{1}{3 |{\bf q} |} \;
\Biggl[ \Pi_3({\bf 0}, {\bf -q}\; ; \Gamma_1 ; 3)  +
\Pi_1( {\bf 0}, {\bf -q}\; ; \Gamma_3 ; 3)  \Biggl]
$$

\vspace*{-0.3cm}
\be
{\cal G}_{C2} = {\cal A} \; \frac{M_\Delta}{{\bf q}^{\;2}} \;
\Pi_1({\bf 0}, {\bf -q}\; ; -i \;\Gamma_1 ; 4)
\ee

\noindent
where ${\cal A}$ is a kinematical factor.

\begin{figure}[h]
\vspace*{-0.4cm}
\tabskip=0pt\halign to\hsize{\hfil#\hfil\tabskip=0pt plus1em&
\hfil#\hfil\cr
\mbox{\includegraphics[height=2.6cm,width=3.5cm]{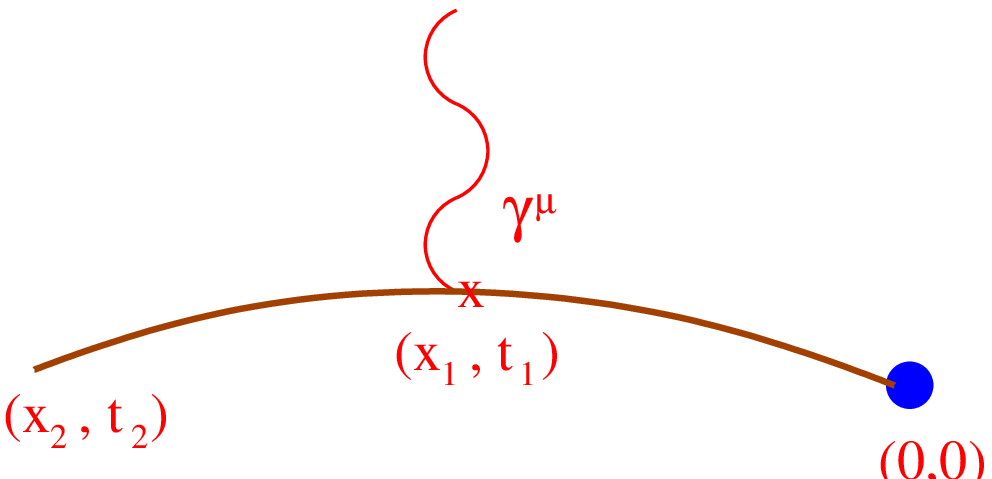}}&
\mbox{\includegraphics[height=2.6cm,width=3.5cm,]{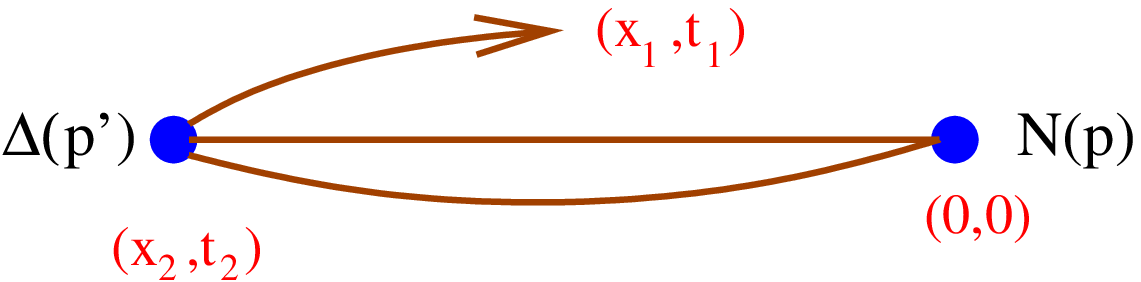}}\cr
(a)&(b)\cr}
\vspace*{-0.75cm}
\caption{(a) Fixed operator and (b) fixed source sequential propagators.}
\label{sequential}
\vspace*{-0.75cm}
\end{figure}

We use
 two methods
 to compute the sequential propagator needed to build the 3-point function:
({\it a}) We evaluate the quark line with the photon insertion,
shown schematically in Fig.~\ref{sequential}(a) by computing
the sequential propagator at fixed momentum transfer ${\bf q}$
and fixed time $t_1$. We look for a plateau by varying the sink-source
 separation
time  $t_2$. The final and initial states can be chosen at
the end.
({\it b}) We evaluate the backward sequential propagator shown schematically
in Fig.~\ref{sequential}(b) by fixing the initial and final states.
$t_2$ is fixed and a plateau is searched for by varying $t_1$. Since the
momentum transfer is specified only at the end, the $\gamma N\rightarrow \Delta$ form factors can
be evaluated at all lattice momenta.

\begin{table}
\caption{}
\small
\label{table:parameters}
\begin{tabular}{|c|c|c|c|} \hline
 \multicolumn{1}{|c|}{$ Q^2$ (GeV$^2$)} &$\kappa$ & $m_\pi/m_\rho$ & Number of confs \\ \hline
\multicolumn{4}{|c|} {Quenched $\beta=6.0$ $16^3\times 32$ ${\bf q}^2 =0.64$ GeV$^2$} \\ \hline
0.64   & 0.1530 & 0.84 & 100 \\
0.64   & 0.1540 & 0.78 & 100 \\
0.64   & 0.1550 & 0.70 & 100 \\ \hline
\multicolumn{4}{|c|} {Quenched $\beta=6.0$ $32^3\times 64$ $ {\bf q}^2=0.64$ GeV$^2$ } \\ \hline
0.64   & 0.1550 & 0.69 & 100 \\ \hline
\multicolumn{4}{|c|} {Quenched $\beta=6.0$ $32^3\times 64$ $ {\bf q}^2=0.16$ GeV$^2$ } \\ \hline
0.16  & 0.1554 & 0.64 & 100 \\
0.15  & 0.1558 & 0.59 & 100  \\
0.13  & 0.1562 & 0.50 & 100   \\ \hline
\multicolumn{4}{|c|} {Unquenched $\beta=5.6$ $16^3\times 32$~\cite{SESAM} $ {\bf q}^2=0.54$ GeV$^2$ }\\ \hline
0.54   & 0.1560  & 0.83 & 196\\
0.54   & 0.1565  & 0.81 & 200  \\
0.54   & 0.1570  & 0.76 & 201  \\
0.54   & 0.1575  & 0.68 & 200  \\ \hline
\end{tabular}
\vspace*{-1cm}
\end{table}

\begin{figure}[h]
\vspace*{-0.5cm}
\mbox{\includegraphics[height=6.5cm,width=7cm]{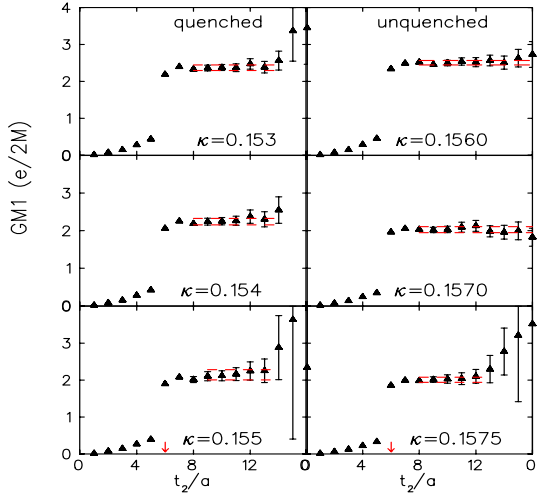}}
\vspace*{-1cm}
\caption{${\cal G}_{M1}$ at  ${\bf q}=(2\pi/16a,0,0)$ for the quenched (left) and for the unquenched (right) theory.
The photon couples to the quark
at $t_1/a=6$ 
as  shown by the arrow.
The lines show the plateau fit range and
bounds obtained by jackknife analysis.}
\vspace*{-0.8cm}
\label{fig:GM1}
\end{figure}

The parameters of our lattices are given in Table~\ref{table:parameters}, where
 we used  the nucleon  mass in the chiral
limit to convert to physical units, and $Q^2=-q^2$ is evaluated in the rest frame of the $
\Delta$.

\begin{figure}[h]
\vspace*{-0.5cm}
\mbox{\includegraphics[height=6.5cm,width=7cm]{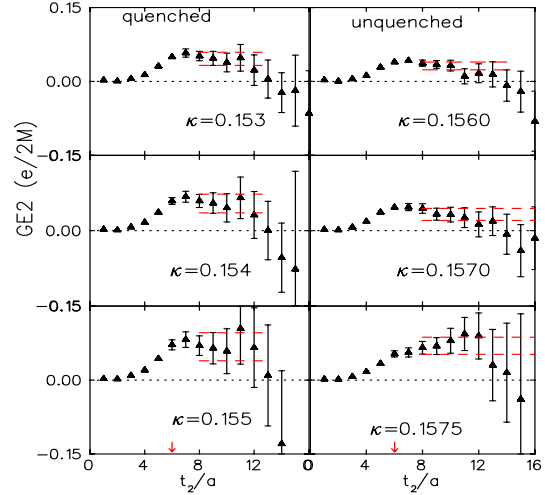}}
\vspace*{-1cm}
\caption{${\cal G}_{E2}$ at  ${\bf q}=(2\pi/16a,0,0)$.
The notation is the same as that of Fig.~\ref{fig:GM1}.}
\vspace*{-0.8cm}
\label{fig:GE2}
\end{figure}

\begin{figure}[h]
\mbox{\includegraphics[height=6cm,width=6cm]{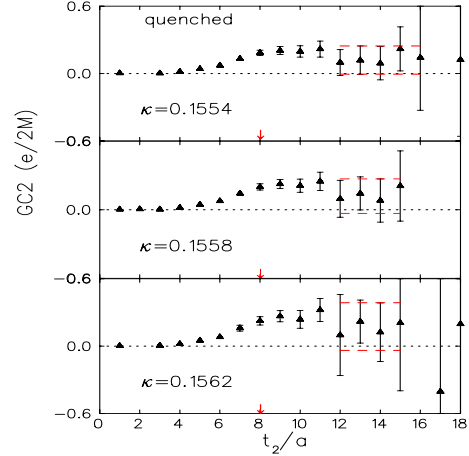}}
\vspace*{-1cm}
\caption{${\cal G}_{C2}$ at ${\bf q}=(2\pi/32a,0,0)$ for the quenched theory.
The photon couples to the quark
at $t_1/a=8$ 
as  shown by the arrow.
The rest of the  notation is the same as that of Fig.~\ref{fig:GM1}.}
\vspace*{-0.8cm}
\label{fig:GC2}
\end{figure}

 We check for finite volume effects
by comparing results
in the quenched theory on
lattices of size $16^3\times 32$ and $32^3\times 64$
at  the same momentum transfer at $\kappa=0.1550$.
Assuming a 1/volume dependence we find that on the small
volumes there is a $(10-15)\%$ correction as compared to the infinite
volume result, whereas
on the large lattice the volume correction  is negligible.
In Figs.~\ref{fig:GM1} and \ref{fig:GE2} we  show quenched and
unquenched results for ${\cal G}_{M1}$ and ${\cal G}_{E2}$ at the same momentum transfer.
Unquenching
tends  to decrease ${\cal G}_{M1}$ and ${\cal G}_{E2}$  but
leaves
 the ratio $R_{EM}=-{\cal G}_{E2}/{\cal G}_{M1}$ 
largely unaffected for the SESAM quark masses
studied in this work, giving values in the range of $-(2-4)\%$.
 The fact that no
increase of  $R_{EM}$ is observed
means that pion contributions to this ratio are small for the  SESAM pion masses.
In Fig.~\ref{fig:GC2} we also show results for
 ${\cal G}_{C2}$ with $\Delta$ static
for the large quenched lattice for which
we obtain the best signal.
Although  ${\cal G}_{C2}$ is within one standard deviation of zero, it is positive
at all $\kappa$-values giving  a negative
ratio $R_{SM} = -|{\bf q}|/2m_\Delta \> {\cal G}_{C2}/{\cal G}_{M1}$ 
in agreement with experiment.

Chiral extrapolation of the results  is done
linearly in the pion mass squared,
since with the nucleon or the $\Delta$ carrying
a finite momentum,  chiral logs are expected to be suppressed.
The values obtained are given in  Table~\ref{table:chiral}
and are in reasonable agreement with the experimental values
$R_{EM}=-2.1 \pm 0.2 \pm 2.0$ at  $Q^2=0.126$~GeV$^2$ and
$R_{EM}=-1.6 \pm 0.4 \pm 0.4$  at  $Q^2=0.52$~GeV$^2$.

\begin{figure}[h]
\mbox{\includegraphics[height=4.9cm,width=7cm]{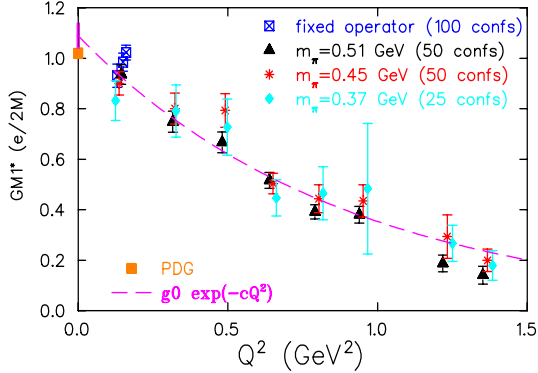}}
\vspace*{-1cm}
\caption{${\cal G}_{M1}^{\star}$ versus $Q^2$
  in the  $\Delta$ rest frame at $\kappa=0.1554$, 0.1558 and 0.1562.
The dashed line is an exponential fit to all the lattice data.
The filled square is the value of ${\cal G}_{M1}^{\star}$
at $Q^2=0.0$ given by the Particle Data Group. }
\vspace*{-0.8cm}
\label{GM1-q2}
\end{figure}

Using the fixed sink method  for the sequential
propagator we  obtain
the $q^2$-dependence of
the form factors.
In Fig.~\ref{GM1-q2} we show 
 preliminary results for  
\be
{\cal G}_{M1}^{\star} \equiv \frac{1}{3}\frac{1}{\sqrt{1+\frac{Q^2}{(m_N+m_\Delta)^2}}} \> {\cal G}_{M1}
\ee
obtained using 50 quenched configurations  at $\kappa=0.1554$
and 0.1558, and 25  at $\kappa=0.1562$.
Here we have used the ratio
\be
R_{\sigma}= \frac{\langle G^{\Delta j^\mu N}_\sigma(t_2,t_1;{\bf p}^{\prime},{\bf p};\Gamma)\rangle}
{\langle \delta_{ij} G_{ij}^{\Delta\Delta}(t_2;{\bf p}^{\prime};\Gamma_4)\rangle
\biggl[\frac{\langle G^{NN}(2t_1;{\bf p};\Gamma_4\rangle)}{
\langle \delta_{ij} G_{ij}^{\Delta\Delta}(2t_1;{\bf p}^{\prime};\Gamma_4)\rangle} \biggr] ^{1/2}}
\label{R-ratio asymmetric}
\ee
to extract the form factors, since using the symmetric combination
given in Eq.~\ref{R-ratio} would require an additional sequential inversion.
At the lowest momentum transfer we confirm that the  
results from Eqs.~\ref{R-ratio} and \ref{R-ratio asymmetric} are
within error bars, as shown by the data at $Q^2=0.13$~GeV$^2$ included in Fig.~\ref{GM1-q2}.
The quark mass dependence is weak especially at the higher momenta
and a global fit to 
the  lattice data may be performed.
The fit 
shows that the lattice data are consistent
with  a simple exponential
dependence on $Q^2$. This fit produces a value at $Q^2=0$  in agreement with the Particle Data Group
result~\cite{PDG}.

\begin{table}
\caption{}
\label{table:chiral}
\small
\vspace*{-0.2cm}
\begin{tabular}{|c|c|c|c|}
\hline
 $Q^2$~GeV$^2$ &  ${\cal G}_{M1}\>(e/2m_N)$ &  ${\cal G}_{E2}\>(e/2m_N)$ & {$R_{EM}\%$} \\ \hline
\multicolumn{4}{|c|}{Quenched QCD} \\ \hline
  0.64 &  1.72(6)  & 0.099(19) & -5.1(1.1) \\
 0.13 &   2.51(6) & 0.104(12) & -4.5(1.4) \\\hline
\multicolumn{4}{|c|}{Unquenched QCD}  \\ \hline
  0.53 &  1.30(4)  & 0.050(31) & -2.8(1.6) \\
\hline
\end{tabular}
\vspace*{-0.8cm}
\end{table}

\section{Conclusions}

Lattice techniques are shown to be suitable   
 for the evaluation of hadron charge and matter density
distributions.
Differences between quenched and unquenched
results are shown to be insignificant, for both the charge and matter 
density distributions.
 The
charge density distribution is, in all cases,
broader than the  matter density. For baryons, the lattice indicates a charge  radius, which is 
$\sim$~20\% larger than the  matter radius.
 The deformation seen in the rho charge distribution 
is absent in the matter
distribution, both in the quenched and the unquenched theory. This
observation suggests a relativistic origin for the deformation
 and  deserves a more extended study, with lighter quarks
and larger volumes.
Nucleon deformation is signaled by a non-zero quadrupole strength in the 
transition $\gamma N \rightarrow  \Delta$.
We have calculated, for the
first time using lattice QCD, the electric quadrupole form factor to high
enough accuracy to exclude a zero value.
The ratio $R_{EM}$  
is evaluated in the 
kinematical regime explored by experiments. The values extracted in the 
chiral limit  are
in good agreement with recent measurements.
Although large statistical and systematic errors prevent an accurate
determination of the Coulomb quadrupole form factor and a zero
value cannot be excluded, our results support a negative value of
the ratio $R_{SM}$ in agreement with experiment.
We have shown preliminary results on the 
$Q^2$ dependence of the magnetic dipole transition form factor
evaluated with $\sim$ 10\% accuracy in the regime explored by Jefferson Lab.
The main conclusion from the analysis of the SESAM lattices~\cite{SESAM}
is that, for pions in the range of 800-500 MeV, no unquenching effects can be
established for $R_{EM}$ within our statistics. We plan
to repeat this calculation for lighter pions to investigate  the pion
cloud contributions on the transition form factors.

\vspace*{0.2cm}

\noindent
{\bf Acknowledgments}: I am grateful to my collaborators
  Ph.~de~Forcrand, Th.~Lippert, H.~Neff, J.~W.~Negele,
K.~Schilling, W. Schroers and A. Tsapalis for their valuable 
contribution  on various
aspects of this work.


\vspace*{-0.1cm}

\begin{thebibliography}{99}

\vspace*{-0.2cm}
%
\bibitem{Isgur} N.~Isgur, G.~Karl and R.~Koniuk,
Phys.\ Rev.\ D {\bf 25} (1982) 2394.
\bibitem{Eisenberg} G. K\"albermann and J. Eisenberg, Phys. Rev. D {\bf 28} (1982) 71; K. Bermuth, D. Drechsel, L. Tiator and J. B. Seaborn, Phys. Rev. D
{\bf 38} (1988) 89.
\bibitem{Buchmann}
A.~J.~Buchmann and E.~M.~Henley,
Phys.\ Rev.\ C {\bf 63} (2001) 015202.
\bibitem{Bates}C.~Mertz {\it et al.},
Phys.\ Rev.\ Lett.\  {\bf 86} (2001) 2963; K. Joo {\it et al.}, 
Phys.\ Rev.\ Lett.\ {\bf 88} (2002) 122001.
\bibitem{Jones} H. F. Jones and M. D. Scadron, Ann. Phys. {\bf 81}, 1 (1973).
\bibitem{SESAM} 
N.~Eicker {\it et al.}, 
Phys.\ Rev.\ D {\bf 59} (1999) 014509.
\bibitem{AFT2} C. Alexandrou, Ph. de Forcrand and A. Tsapalis, Phys. Rev. D {\bf 68} (2003) 074504; hep-lat/0309064.
\bibitem{AFT} C. Alexandrou, Ph. de Forcrand and A. Tsapalis, Phys. Rev. D {\bf66} (2002) 094503.
\bibitem{Negele} M. Burkardt, J. M. Grandy and J. W. Negele, 
Ann. Phys.\  {\bf 238} (1995) 441.
\bibitem{panic02} C. Alexandrou, Ph. de Forcrand and A. Tsapalis, Nucl. Phys. 
{\bf A721} (2003) 907; Nucl. Phys. B (Proc. Suppl.) {\bf 119} (2003) 422.
\bibitem{Gupta} R. Gupta, D. Daniel and J. Grandy, Phys. Rev. D {\bf 48} (1993) 3330.
\bibitem{Leinweber}
D. B. Leinweber, T. Draper and R. M. Woloshyn,
Phys.\ Rev.\ D {\bf 48} (1993) 2230.
\bibitem{All} C. Alexandrou {\it et al.}, 
Nucl. Phys. B (Proc. Suppl.) {\bf 119} (2003) 213; hep-lat/0307018; hep-lat/0309041.
\bibitem{PDG} D. E. Groom {\it et al.}, European Phys. J. C15 (2000) 1.
\end{thebibliography}
\end{document}